\documentclass[conference]{IEEEtran}
\IEEEoverridecommandlockouts
\usepackage{cite}
\usepackage{amsmath,amssymb,amsfonts}
\usepackage{algorithmic}
\usepackage{graphicx}
\usepackage{textcomp}
\usepackage{xcolor}
\usepackage{hyperref}
\def\BibTeX{{\rm B\kern-.05em{\sc i\kern-.025em b}\kern-.08em
    T\kern-.1667em\lower.7ex\hbox{E}\kern-.125emX}}

\begin{document}

\title{BEAR: BGP Event Analysis and Reporting}

\author{\IEEEauthorblockN{1\textsuperscript{st} Hanqing Li}
\IEEEauthorblockA{\textit{Engineering Sciences and Applied Mathematics Department} \\
\textit{Northwestern University}\\
Evanston, USA \\
hanqingli2025@u.northwestern.edu}
\and
\IEEEauthorblockN{2\textsuperscript{nd} Melania Fedeli}
\IEEEauthorblockA{\textit{Amazon Web Services (AWS)} \\
Dublin, Ireland \\
melaniaf@amazon.com}
\and
\IEEEauthorblockN{3\textsuperscript{rd} Vinay Kolar}
\IEEEauthorblockA{\textit{Amazon Web Services (AWS)} \\
Cupertino, USA \\
vinkolar@amazon.com}
\and
\IEEEauthorblockN{4\textsuperscript{th} Diego Klabjan}
\IEEEauthorblockA{\textit{Industrial Engineering and Management Sciences Department} \\
\textit{Northwestern University}\\
Evanston, USA \\
d-klabjan@northwestern.edu}
}

\maketitle

\begin{abstract}
The Internet comprises of interconnected, independently managed Autonomous Systems (AS) that rely on the Border Gateway Protocol (BGP) for inter-domain routing. BGP anomalies—such as route leaks and hijacks—can divert traffic through unauthorized or inefficient paths, jeopardizing network reliability and security. Although existing rule-based and machine learning methods can detect these anomalies using structured metrics, they still require experts with in-depth BGP knowledge of, for example, AS relationships and historical incidents, to interpret events and propose remediation. In this paper, we introduce BEAR (BGP Event Analysis and Reporting), a novel framework that leverages large language models (LLMs) to automatically generate comprehensive reports explaining detected BGP anomaly events. BEAR employs a multi-step reasoning process that translates tabular BGP data into detailed textual narratives, enhancing interpretability and analytical precision. To address the limited availability of publicly documented BGP anomalies, we also present a synthetic data generation framework powered by LLMs. Evaluations on both real and synthetic datasets demonstrate that BEAR achieves 100\% accuracy, outperforming Chain-of-Thought and in-context learning baselines. This work pioneers an automated approach for explaining BGP anomaly events, offering valuable operational insights for network management.
\end{abstract}


\section{Introduction}
\label{sec:intro}
The Border Gateway Protocol (BGP) is the principal inter-domain routing protocol that facilitates data exchange across the Internet by enabling autonomous systems (ASes) to disseminate network reachability information \cite{rekhter2006border}. As the backbone of Internet connectivity, BGP's proper functioning is critical for maintaining global network stability and performance \cite{labovitz1998internet}. 

\begin{figure}
    \centering
    \includegraphics[width=\columnwidth]{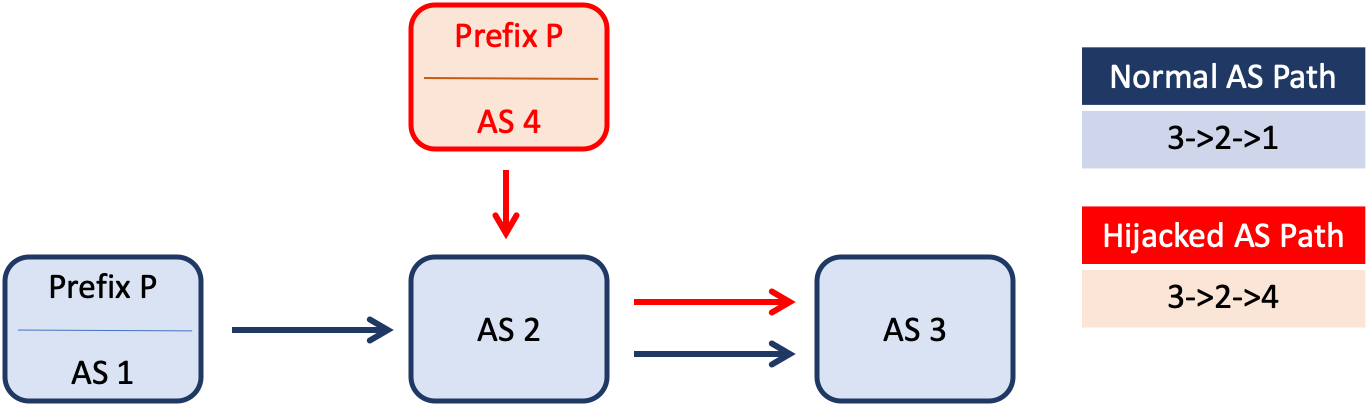}
    \caption{BGP hijack example.}
    \label{fig:hijack}
\end{figure}

\begin{figure}
    \centering
    \includegraphics[width=\columnwidth]{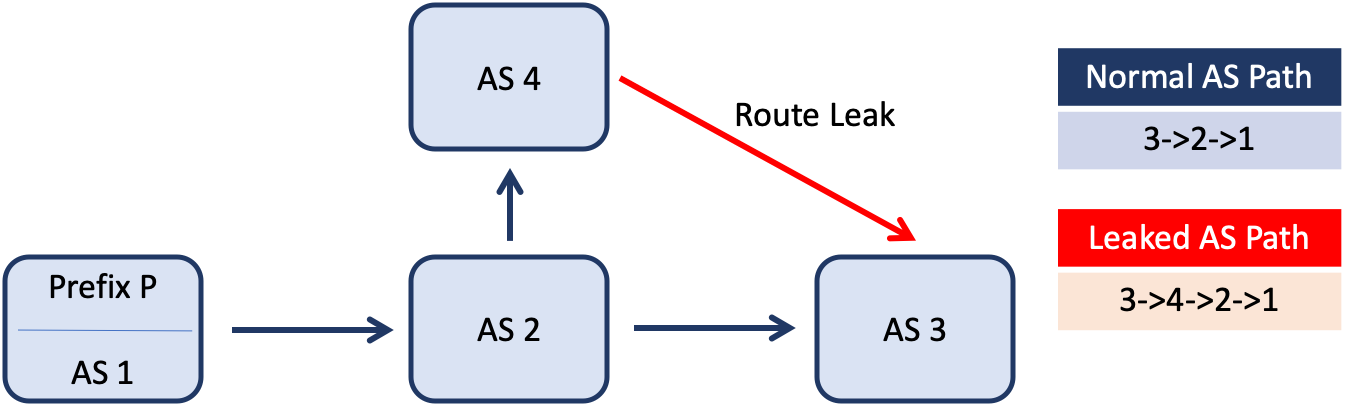}
    \caption{BGP route leak example.}
    \label{fig:leak}
\end{figure}

However, BGP's inherent lack of built-in security measures renders it susceptible to various anomalies, notably BGP hijacking and route leaks\cite{al2015detecting}. BGP hijacking involves an AS illegitimately announcing IP prefixes it does not own, thereby diverting or intercepting traffic intended for the legitimate IP address holder\cite{al2016bgp} (see Figure \ref{fig:hijack}). This can lead to significant disruptions, including loss of data confidentiality and integrity. Route leaks occur when an AS improperly announces received BGP routes to unintended parties, often due to misconfigurations or policy violations\cite{al2016bgp} (see Figure \ref{fig:leak}). Such leaks can cause suboptimal routing, traffic congestion, and widespread service outages, undermining the reliability of Internet services. The ramifications of these BGP vulnerabilities are profound, potentially leading to large-scale Internet disruptions, economic losses, and compromised security. Therefore, detecting and mitigating BGP anomalies are imperative to uphold the resilience and trustworthiness of the global Internet infrastructure.

While existing methods can detect the occurrence of BGP anomalies \cite{scott2025bgp,scott2024matrix,xu2020bgp,hoarau2021suitability,sunita2024optimal,dong2021isp,moriano2021using,latif2022unveiling,peng2022multi,fezeu2020anomalous,huang2024realtime}, a comprehensive understanding of these events is essential for effective mitigation and prevention. Detailed insights into the event type, affected ASes, pre- and post-event path changes, and identification of the malicious or misconfigured AS are crucial for network operators and security professionals. Such in-depth analysis enables targeted responses, minimizes disruption, and enhances the overall security posture of the Internet's routing infrastructure. Consequently, it is imperative to develop methods that can automatically generate comprehensive reports upon the detection of an anomaly. We refer to this challenge as BGP anomaly event explanation and define it formally in Section \ref{subsec:pro_def}.

Large Language Models (LLMs) have demonstrated remarkable proficiency in automatic text generation and complex reasoning tasks \cite{zhao2023survey}. While structured machine learning models—such as graph-based or other structured approaches—excel at detecting anomalies using predefined metrics \cite{al2016bgp}, they are inherently limited to the information explicitly provided to them and lack the broader world knowledge required to interpret these anomalies. For example, although structured models can flag deviations in BGP metrics, they cannot readily account for historical patterns of regional attacks or subtle indicators of sophisticated routing exploits. In contrast, LLMs are imbued with extensive domain and world knowledge, including detailed insights into BGP operations and known vulnerabilities, enabling them to generate nuanced, context-rich explanations. However, LLMs are not ideally suited for the direct extraction of precise metric anomalies from large-scale data, due to their training on textual rather than structured numerical data \cite{zhou2025can}. Thus, while we use structured models to detect anomalies, our novelty lies in employing LLMs to interpret and explain these anomalies in detail.

In this paper, we propose a novel framework, BEAR (BGP Event Analysis and Reporting), designed to leverage an LLM for generating comprehensive reports that explain detected BGP anomaly events. The framework begins by extracting relevant BGP data associated with an anomaly from an online BGP database using the provided timestamp and IP prefix. Recognizing the strengths of LLMs in handling textual data over tabular formats, we adopt a multi-step reasoning approach: transforming BGP data into text before utilizing the LLM for data analysis and report generation. To enhance the accuracy of the generated reports, we incorporate self-consistency mechanisms and prompt augmentation techniques. We also cover robustness by studying situations with a subset of collectors being online. As a side effect, this also reduces the number of LLM tokens being used. This study focuses on two primary types of BGP anomalies, as categorized in the Zhao et al.'s taxonomy \cite{zhao2023survey}: (1) direct intended anomalies (e.g., BGP hijacks) and (2) direct unintended anomalies (e.g., BGP route leaks). These two categories are prioritized because they represent the most significant types of BGP anomalies and can be effectively analyzed using available public data. In contrast, the other two categories—indirect anomalies and link failures—often require additional external private datasets for comprehensive analysis. As such, this work limits its scope to BGP hijacks and route leaks, leaving the exploration of indirect anomalies and link failures for future research.

Moreover, we introduce the first approach to generate synthetic BGP anomaly event data, addressing the scarcity of fully documented BGP anomaly events. To create high-quality synthetic events, we leverage an LLM to produce specific details of an anomaly, such as the timestamp, victim IP prefix, event type, hijacker or route leaker, AS path after the event, and detection rate. Using the generated details, we extract relevant BGP data from an existing BGP dataset based on the timestamp and victim IP prefix. This data is then modified according to the LLM-specified details, creating synthetic BGP data that simulates the occurrence of an anomaly event. This method produces high-quality synthetic events. We evaluate BEAR on both real-world BGP anomaly events, including many recent incidents, and synthetic events. All generated reports are assessed by BGP experts for correctness. BEAR achieves 100\% accuracy on both real and synthetic datasets.

Our contributions are listed next.
\begin{itemize}
    \item \textbf{Problem Definition and Exploration:} This work is the first to introduce and investigate the problem of BGP anomaly event explanation, aiming to provide insights from both structured data and LLM global knowledge into events by analyzing BGP data both before and after the anomaly event.
    \item \textbf{Novel Multi-Step Reasoning Approach:} The paper proposes a novel multi-step reasoning approach by transforming large-scale tabular BGP data into textual descriptions for enhanced interpretability and performing data analysis and reasoning on the text description instead of the tabular data for high reasoning accuracy. By leveraging prompt augmentation and self-consistency mechanisms, the approach achieves 100\% accuracy and generates detailed BGP anomaly event reports, effectively identifying missing key elements such as event types, relevant ASes, and comprehensive explanations.
    \item \textbf{Robustness of Availability of Collectors}: In real world, collectors can be down. We study such an impact by creating settings with limited availabilities of collectors. With respect to LLMs, this accommodates the use of LLMs with a limited number of input tokens. Leveraging fewer collectors also results in a faster explanation of the events and a quicker mitigation and remediation, thus reducing network impacts.
    \item \textbf{Synthetic Data Generation Framework:} In this research, an innovative framework is presented that utilizes LLMs to synthesize high-quality synthetic BGP anomaly event datasets. This framework generates critical features such as event types, relevant ASes, target IPs, timestamps, and AS paths before and after the events, marking the first method designed to synthesize BGP anomaly events.
\end{itemize}
All codes are available in \url{https://github.com/hanklee97121/BEAR_BGP_EVENT_ANALYSIS}.
\section{Related Research}
\label{sec:related_research}
This section reviews related literature on BGP anomaly detection and the applications of LLMs in the BGP domain.

\subsection{Background}
Each AS advertises the IP prefixes it controls—collections of IP addresses that define reachable networks—which together form a global routing table\cite{schlamp2016cair}. In addition to merely announcing these prefixes, an AS advertises complete routes that include essential attributes such as the next hop and the AS path\cite{schlamp2016cair}. The AS path is a sequential list of AS numbers that a route advertisement has traversed; each time a router forwards the announcement, it prepends its own AS number to the path, thereby creating a path to reach an IP prefix that will be filtered by policy-based routing rules by following the AS to decide whether to take this path and advertise it or not\cite{al2016bgp}. BGP was originally designed to prioritize functionality and scalability rather than security, so it lacks built-in mechanisms to authenticate the source of route announcements\cite{buhler2023oscilloscope}. Because of this trust-based design, any AS can announce routes for IP prefixes regardless of ownership, leaving BGP vulnerable to both configuration errors and deliberate attacks such as hijacks and route leaks.

To study and mitigate these anomalies, researchers rely on open-source collector frameworks that aggregate BGP data from routers worldwide. RIPE RIS and RouteViews are two prominent projects that collect global BGP data by acting as route collectors\cite{routeviews2013university, ripe}. Both projects establish BGP peering sessions with numerous volunteer networks to receive routing information in real time. RIPE RIS deploys a network of Remote Route Collectors at strategic locations—often at major Internet Exchange Points—where they peer with participating networks. When an AS peers with an RIS collector, it sends a complete snapshot of its Routing Information Base (RIB dump) as well as continuous incremental update messages that reflect changes in its routing table. These data are recorded in the standardized MRT (Multi-threaded Routing Toolkit) format, making them available for public analysis and research. Similarly, the RouteViews project collects BGP data by establishing voluntary BGP sessions with a wide range of networks.

\subsection{BGP anomaly detection}
\label{sec:anomaly_detection}
Over the years, various techniques have been developed to detect BGP anomalies\cite{al2016bgp}. Traditional methods primarily rely on statistical analyses and rule-based systems to identify irregularities in BGP update messages\cite{al2016bgp}. One category of approaches involves time-series analysis of observed BGP features. Labovitz et al.\cite{labovitz1998internet} introduced one of the earliest methods, leveraging Fast Fourier Transform to analyze routing update rates. Subsequent work by Mai et al.\cite{mai2008detecting} and Prakash et al.\cite{prakash2009bgp} employed wavelet transforms to analyze multiple BGP features for anomaly detection. Al-Musawi et al.\cite{al2015detecting} proposed Recurrence Quantification Analysis, a non-linear dynamics technique, to detect BGP instability by examining BGP volume and the average AS-path length.

Another category involves traditional machine learning techniques, such as decision trees, Naive Bayes, and support vector machines \cite{li2005internet,de2011anomaly,al2012machine,lutu2014separating}. For instance, Al-Rousan and Trajkovic \cite{al2012machine} applied both support vector machines and Hidden Markov Models to 37 selected features to detect BGP anomalies. Lutu et al.\cite{lutu2014separating} used the windowing algorithm on nine selected BGP features for anomaly detection.

Statistical pattern recognition techniques have also been explored, including Principal Component Analysis \cite{huang2007diagnosing}, generalized likelihood ratio tests \cite{deshpande2009online}, higher-order path analysis \cite{ganiz2006detection}, and Z-score calculations \cite{theodoridis2012novel}. Other methods involve brute-force comparisons of historical BGP data \cite{karlin2006pretty,lad2006phas,haeberlen2009netreview,shi2012detecting} or assessing prefix reachability \cite{zheng2007light,hu2007accurate,tahara2008method,zhang2008ispy}.

Recent advancements integrate neural networks into anomaly detection, utilizing techniques such as LSTMs \cite{dong2021isp,moriano2021using} and Graph Neural Networks \cite{latif2022unveiling,peng2022multi}. Fezeu et al.\cite{fezeu2020anomalous}, Huang et al.\cite{huang2024realtime}, and others\cite{xu2020bgp,hoarau2021suitability,sunita2024optimal} further demonstrate the potential of neural networks for BGP anomaly detection. Scott et al.\cite{scott2024matrix} applied the Matrix Profile, a time-series mining approach, to detect BGP anomalies across various event categories. More recently, Scott et al.\cite{scott2025bgp} proposed the multidimensional recurrence quantification analysis, treating BGP anomaly detection as a group dynamics problem by analyzing interactions and temporal relationships among multiple ASes.

Unlike these methods, which primarily focus on detecting BGP anomaly events, our approach addresses the subsequent step: analyzing the detected anomalies and generating comprehensive reports to explain the events.

\subsection{LLM in BGP}
With the advancement of LLMs and their powerful capabilities in text understanding and generation, these models have been applied to various domains, including addressing BGP-related challenges. Mondal et al.\cite{mondal2023llms} explored using LLMs for automated router configuration synthesis, including BGP policy implementation. Kan et al.\cite{kan2024mobile} developed a fine-tuned LLM for network analysis, including IP routing analyses with an emphasis on BGP paths. Palmero et al.\cite{palmero2024providing} introduced an AI-powered assistant leveraging LLMs for tasks such as BGP configuration and anomaly diagnoses.

However, none of these methods address the specific challenge of analyzing BGP anomaly events and generating comprehensive explanatory reports. Our approach bridges this gap, introducing the first method explicitly designed for BGP anomaly event explanation.

\section{Methodology}
\label{sec:method}
This section outlines the methodology for generating the BGP anomaly event report. First, we formally define the BGP anomaly event explanation problem. Next, we describe the process of retrieving the relevant data for an event analysis. Finally, we present the framework for producing high-quality BGP anomaly event reports.

\subsection{Problem Definition}
\label{subsec:pro_def}
Before introducing the method, we formally define the problem of BGP anomaly event explanation. A BGP anomaly event $E$ is given by an associated IP prefix $ip$ and a specific start time $t$. We assume this event is currently detected by any of the BGP anomaly detection algorithms such as the ones explained in Section \ref{sec:anomaly_detection}. Additionally, we have access to the BGP dataset $D_{BGP}$, which contains BGP update messages and routing information over time. The objective is to develop a function $f$ that processes this information to generate a natural language report $R$. The report should explain the details of the event $E$ and identify key unknown features, such as the event type and relevant ASes.

\subsection{Data Retrieval}
\label{subsec:data_ret}
\begin{figure}
    \centering
    \includegraphics[width=0.9\linewidth]{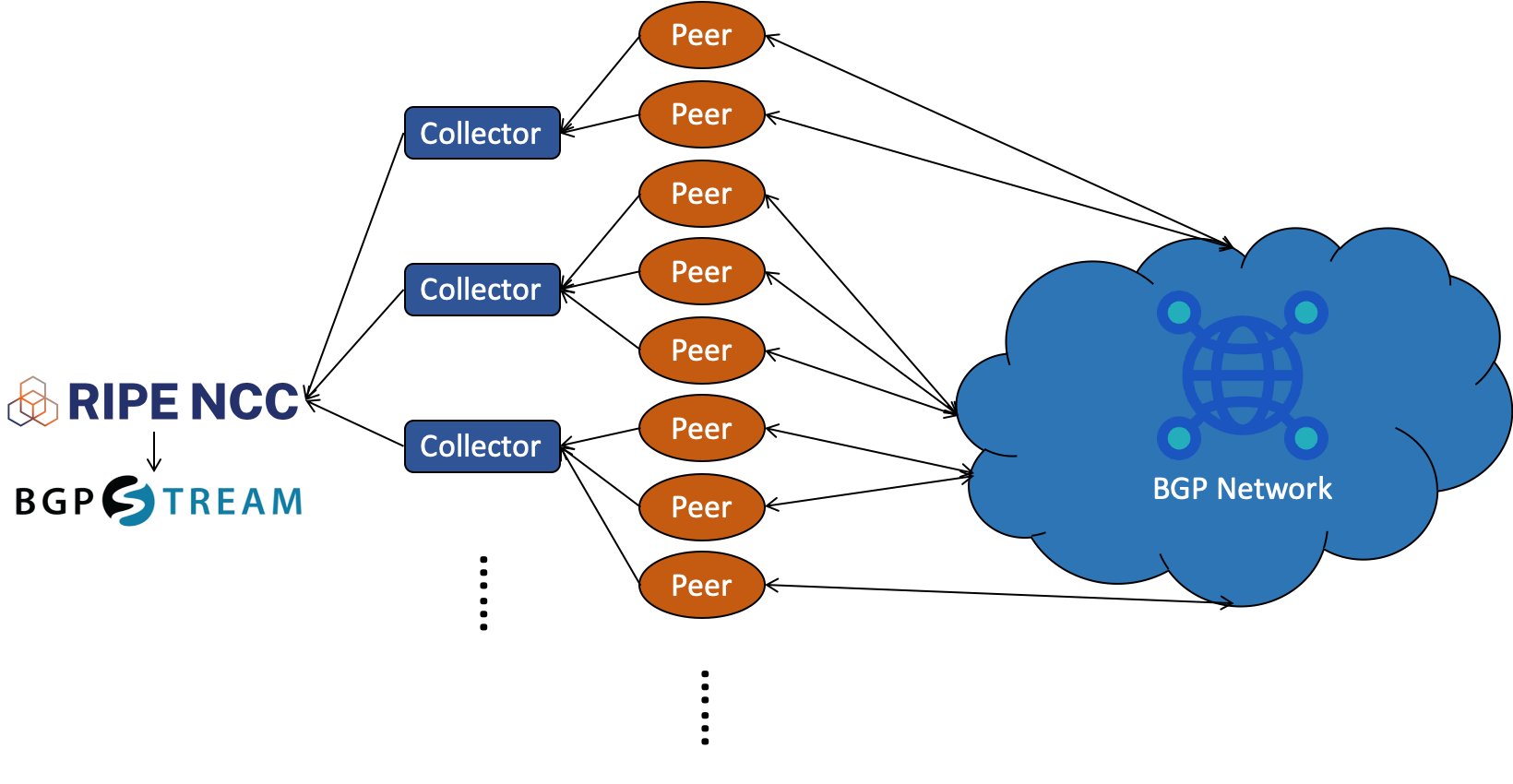}
    \caption{Overview of how BGP messages and routing information are collected.}
    \label{fig:data_retrieve}
\end{figure}

\begin{figure}
    \centering
    \includegraphics[width=0.9\linewidth]{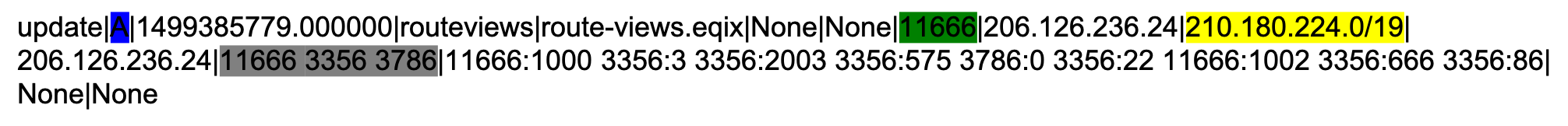}
    \caption{An example of a BGP update message.}
    \label{fig:bgp_message}
\end{figure}
After obtaining the target IP prefix $ip$ and the specific start time $t$, we utilize the open-source software framework BGPStream~\cite{orsini16bgpstream} to retrieve relevant AS path information from the BGP dataset $D_{BGP}$. Specifically, we extract AS paths from BGP update messages and routing information base (RIB) records to construct three datasets: historical AS paths, AS paths before the event, and AS paths after the event. 

BGPStream accesses BGP messages and RIB records collected by RIPE Routing Information Service (RIS)~\cite{ripe}, a routing data collection platform that gathers BGP data through collectors. As illustrated in Figure \ref{fig:data_retrieve}, each collector connects to multiple peers, representing real ASes in the BGP network, and collects BGP update messages sent by these peers every five minutes. Additionally, collectors extract RIBs (i.e., all AS paths from a peer to all accessible IP prefixes) at eight-hour intervals.

Figure \ref{fig:bgp_message} illustrates an example BGP message. The AS path (highlighted in grey) represents the route taken by an AS (highlighted in green) to reach an IP prefix (highlighted in yellow). The letter ``A" (highlighted in blue) indicates an announcement of an AS path, while the letter ``W" denotes a withdrawal message, signaling the removal of an existing AS path.

An RIB record captures a snapshot of the BGP routing table at a specific moment, providing a comprehensive view of all active AS paths known to an AS at that time. The RIPE RIS collectors generate these snapshots every eight hours, collecting RIB records from all their peers. The format of an RIB record is similar to that of a BGP message, with the key distinction being the identifier letter ``R" (highlighted in blue), which denotes an RIB record.

To build the datasets we proceed as follows.
\begin{enumerate}
    \item Historical Data ($D_{history}$): We extract RIB records from all RIPE RIS collectors, capturing all AS paths to $ip$. To ensure $D_{history}$ does not include information about the anomaly event $E$, we use RIB records from at least eight hours prior to the event’s start time $t$. Given that RIB records are collected every eight hours, the timestamp for $D_{history}$ is determined as $\lfloor\frac{t-8h}{8h}\rfloor\times 8h$. Since each AS maintains a single AS path to a prefix at a time, we store one AS path for each peer directly connected to a RIPE RIS collector in $D_{history}$.
    \item Data Before the Event ($D_{before}$): Building on $D_{history}$, we incorporate all BGP update messages from the timestamp of $D_{history}$ up to five minutes before $t$. Each BGP update message modifies the AS paths in $D_{history}$ to reflect the state immediately before the event. Specifically, if a BGP update message is an annoucement of a new AS path from peer $p$ to $ip$, we replace the old one from $p$ in $D_{history}$ with this new AS path. If a BGP update message is a withdrawl of an existing AS path from peer $p$ to $ip$, we delete the AS path from $p$ in $D_{history}$. Starting from $D_{history}$, we incorporate all BGP update messages recorded between the timestamp of $D_{history}$ and five minutes before $t$. Each update message modifies $D_{history}$ to reflect the network state immediately before the event. If an update message announces a new AS path from peer $p$ to $ip$, we replace the corresponding path in $D_{history}$. If the update message withdraws an AS path from peer $p$ to $ip$, the path is removed from $D_{history}$.
    \item Data After the Event ($D_{after}$): $D_{before}$ is further updated using BGP update messages collected between five minutes before $t$ and five minutes after $t$, or up to one second before the event ends if the end time is earlier than $t+5m$. The update process follows the same procedure as in $D_{before}$, ensuring $D_{after}$ reflects the network state after the anomaly event starts.
\end{enumerate}

To account for sub-prefix anomaly scenarios, we also collect RIB records and BGP update messages for IP prefixes that are more or less specific than $ip$, updating $D_{history}$, $D_{before}$, and $D_{after}$ accordingly. All three datasets are stored as structured tabular data in the JSON format, following the structure ``\{collector:\{peer:[AS path]\}\}.'' We refer to these datasets collectively as the BGP data.

\subsection{Data Analysis and Report Generation}
\label{subsec:report_gen}
\begin{figure}
    \centering
    \includegraphics[width=0.9\linewidth]{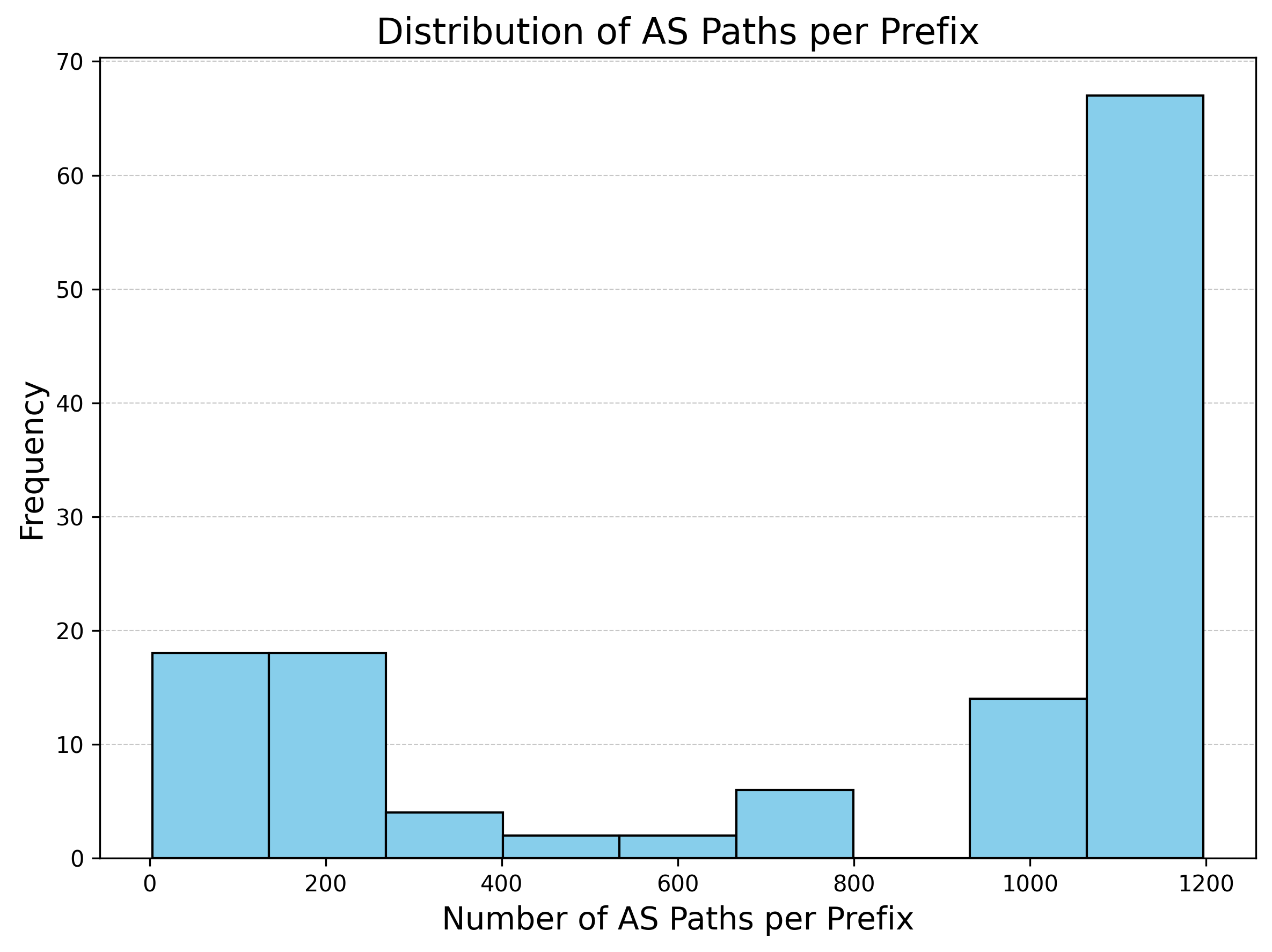}
    \caption{Number of AS path per IP prefix.}
    \label{fig:path_prefix}
\end{figure}
\begin{figure}
    \centering
    \includegraphics[width=0.9\linewidth]{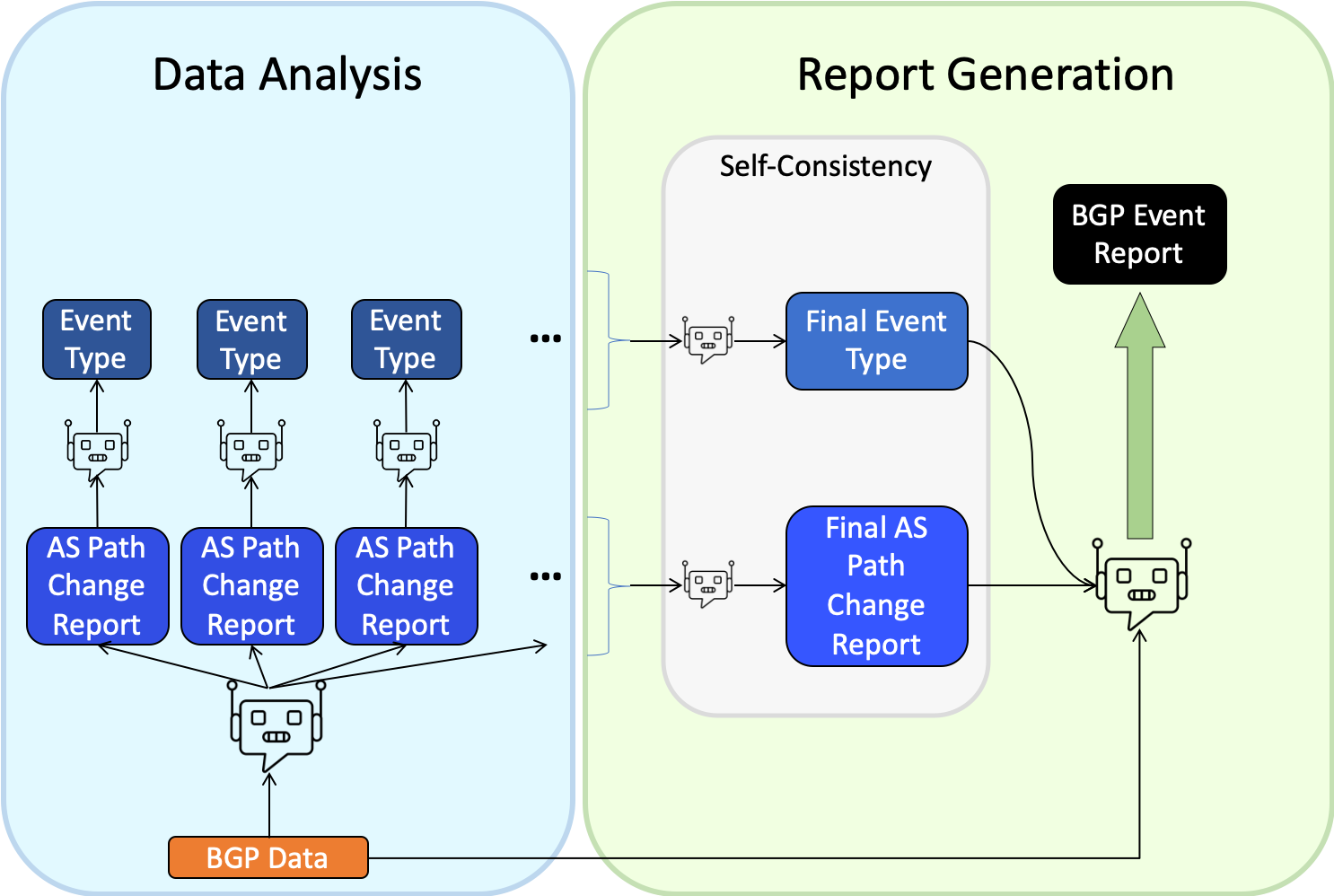}
    \caption{Overview of the framework that generates report for BGP anomaly events}
    \label{fig:framework_flow}
\end{figure}
As illustrated in Figure \ref{fig:path_prefix}, most IP prefixes are associated with between 1,000 and 1,200 recorded AS paths, and each event involves, on average, 2 to 3 IP prefixes in the BGP data. Given the scale of the dataset—which comprises of up to 1,545 peers, with most events recording between 2,000 and 3,600 AS paths—directly analyzing these three datasets to produce an accurate explanation presents a significant challenge for LLMs. To address this, we employ a multi-step reasoning framework that systematically guides the LLM through the report generation process, incorporating prompt engineering, in-context learning, and a self-consistency mechanism to enhance accuracy and reliability.

The method begins with a data analysis phase designed to transform tabular BGP data into textual descriptions, as illustrated in Figure \ref{fig:framework_flow}. The LLM is first provided with the BGP data in a tabular format and prompted to describe changes in AS paths by comparing $D_{after}$ with $D_{before}$, using $D_{history}$ as a reference. The prompt is carefully structured to decompose the reasoning process, directing the LLM to analyze AS path changes systematically. The LLM is required to answer the following key questions.
\begin{itemize}
    \item Does the existing path from each peer to the target IP prefix change?
    \item If it does, does the last AS (destination) change or not?
    \item Is there any new AS path to a new sub-prefix introduced?
    \item If there is, compare it to the existing path with the same peer, is there any difference?
    \item Does the last AS (destination) change or not?
\end{itemize}
These targeted questions guide the LLM to extract meaningful patterns from the BGP data and produce structured responses. In this manner, the tabular BGP data is transformed into descriptive text containing meaningful information, which is then utilized by the LLM to determine the event type and compose the report.

Despite these refinements, we observe that the LLM occasionally misidentifies the destination AS in an AS path. For example, in the AS path [4608, 1221, 4637, 15169], the LLM may incorrectly infer AS4637 as the destination instead of the correct answer, AS15169. To mitigate this, we employ few-shot in-context learning, providing the LLM with examples of AS paths alongside their correct destination AS, ensuring it learns to accurately interpret AS paths.

Next, the LLM is tasked with classifying the anomaly type—determining whether the event is a BGP hijack or a BGP route leak based on observed AS path changes. The prompt explicitly describes the distinguishing characteristics of these anomalies, ensuring the model applies the correct classification criteria. Additionally, we emphasize in the prompt that the consequences of these anomalies may be reflected in a single AS path or a sub-prefix, ensuring that the model accounts for sub-prefix anomalies in its analysis. This emphasis also reinforces the importance of detecting even subtle differences between $D_{before}$ and $D_{after}$, enabling the LLM to identify minor but crucial changes that may indicate an anomaly.

To improve robustness, we repeat this data analysis and classification processes identically $N$ times, generating $N$ AS path change reports and $N$ anomaly classifications. They differ due to the stochastic generation process of LLMs. A self-consistency mechanism is then applied, where we ask the LLM to select the most frequently occurring anomaly classification across all iterations. Then, instead of directly choosing one of the generated AS path change reports, we prompt the LLM to synthesize the final AS path change report that aligns with the majority of the previously generated reports, ensuring consistency and coherence. This approach mitigates variability in the LLM’s outputs and reinforces correctness by prioritizing the most consistent and frequently observed description.

Finally, we provide the BGP data, the final AS path change report, and the final anomaly classification as inputs to the LLM, prompting it to generate a comprehensive BGP event report. The crafted prompt ensures that the report includes key details such as the event type and affected ASes. By combining multi-step reasoning, prompt engineering, and self-consistency, BEAR produces highly accurate and interpretable reports that effectively explain BGP anomaly events.

\section{Dataset}
\label{sec:dataset}
We gather information on 10 well-documented BGP anomaly events from online sources, recording their event type, relevant ASes, target IP, start time, end time, event name, and a link to the event description. The corresponding BGP data for each event is retrieved as outlined in Section \ref{subsec:data_ret}.

Given the detailed online descriptions of these events, many of which occurred prior to 2023, it is possible that the LLM used in our experiments has been trained on these reports. To ensure that the LLM generates explanations by reasoning and analyzing data rather than retrieving reports from its memory, we modified the BGP data for these 10 events. Specifically, we replaced the AS numbers with random values and the timestamps with random timestamps, creating 10 additional samples. These modified events retain the same structural characteristics as the original events but eliminate identifying features that could trigger memory-based retrieval by the LLM. The resulting dataset consists of 20 samples, encompassing both the original events and their anonymized counterparts, which we refer to collectively as the ``real events.''

\subsection{Synthetic BGP Event}
\label{sec:data_gen}
\begin{figure}
    \centering
    \includegraphics[width=0.7\linewidth]{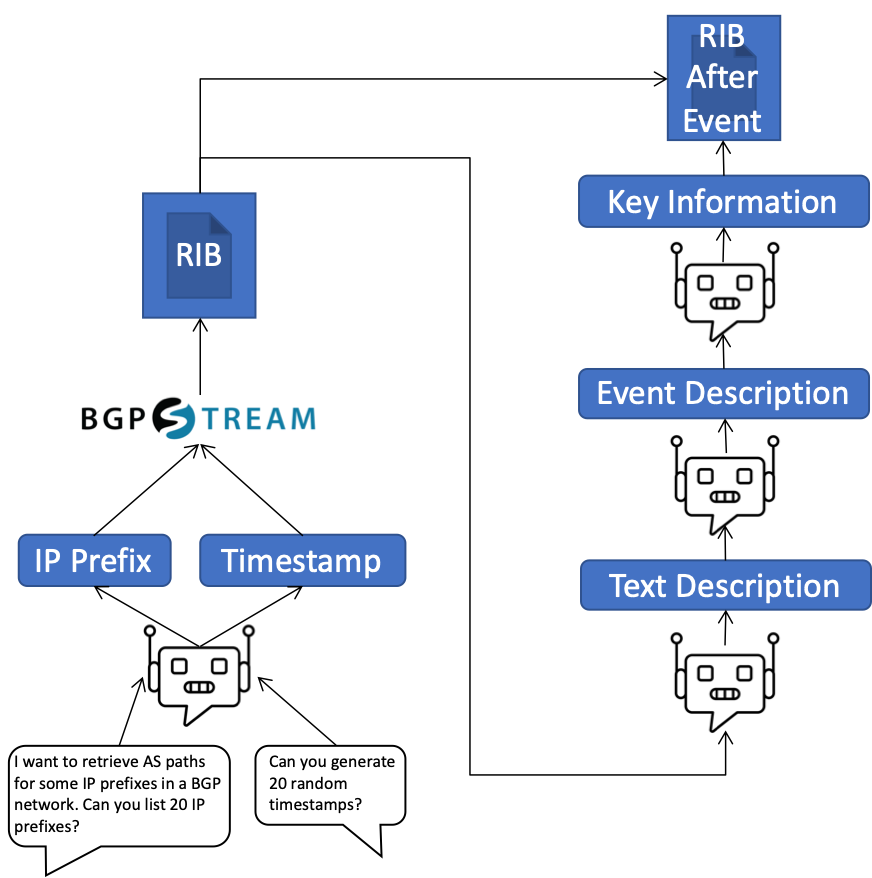}
    \caption{Workflow of synthetic BGP event generation.}
    \label{fig:data_gen}
\end{figure}
Due to the scarcity of public well-documented BGP anomaly event, we only have 20 real events in our dataset to test the performance of the method. To get more samples to evaluate the performance of the method, we develop a framework to generate synthetic BGP events. As illustrated in Figure \ref{fig:data_gen}, we generate a BGP event by generating BGP data ($D_{history}$, $D_{before}$, $D_{after}$) as well as the target IP prefix and timestamp with the aid of an LLM. 

The complete workflow is illustrated in Figure \ref{fig:data_gen}. The main idea is to retrieve real $D_{history}$ and $D_{before}$ from $D_{BGP}$ and to generate a synthetic $D_{after}$ that simulates the impact of a hypothetical BGP anomaly event using an LLM. We begin by prompting the LLM to randomly generate a valid IP prefix and timestamp, which are used to retrieve routing information via BGPStream, forming $D_{history}$ and $D_{before}$. To ensure that these datasets are unaffected by any anomaly event, we verify the consistency of the AS paths between them. The $D_{before}$ dataset is then fed into the LLM, which is asked to produce a textual description, including patterns in the AS paths and examples of specific AS paths.

Next, the LLM is tasked with generating a detailed hypothetical BGP anomaly event description based on the provided AS path patterns and examples. Initially, we ask the LLM to randomly select the event type. Thereafter, it is specifically instructed to produce a description that incorporates key details—namely, the event type, any applicable sub-prefix, the identity of the hijacker or leaker, representative examples of affected AS paths, and the percentage of peers detecting the event. All prompts are provided in a zero-shot manner with detailed instructions.

Using the generated event description, we extract the key information and modify $D_{before}$ to create $D_{after}$. If the event involves a sub-prefix hijack or route leak, all AS paths in $D_{before}$ for the target IP prefix are duplicated and applied to the sub-prefix, after which subsequent edits are made to the sub-prefix data. Peers are sampled based on the percentage of detection, and their AS paths to the target IP prefix are replaced with randomly selected examples provided by the LLM. For AS path replacement, in BGP hijack events, the first AS is set as the peer, the last AS is the hijacker, and the rest is modified based on the selected example. In BGP route leak events, the suffix of the AS path is replaced with the sample AS path provided by the LLM, starting with the leaker and ending with the correct destination AS.

The LLM generates multiple AS path samples for hijack events, while for route leak events, a single AS path sample suffices since the path from the leaker to the valid destination is unique. After these modifications, $D_{after}$ is finalized, completing the generation of one synthetic BGP anomaly event.

We generate 34 synthetic BGP anomaly events, evenly divided into 17 BGP hijacks and 17 BGP route leaks, to assess our method. This dataset is referred to as the ``synthetic'' dataset.

\section{Experiment Results}
\begin{figure}
    \centering
    \includegraphics[width=0.7\linewidth]{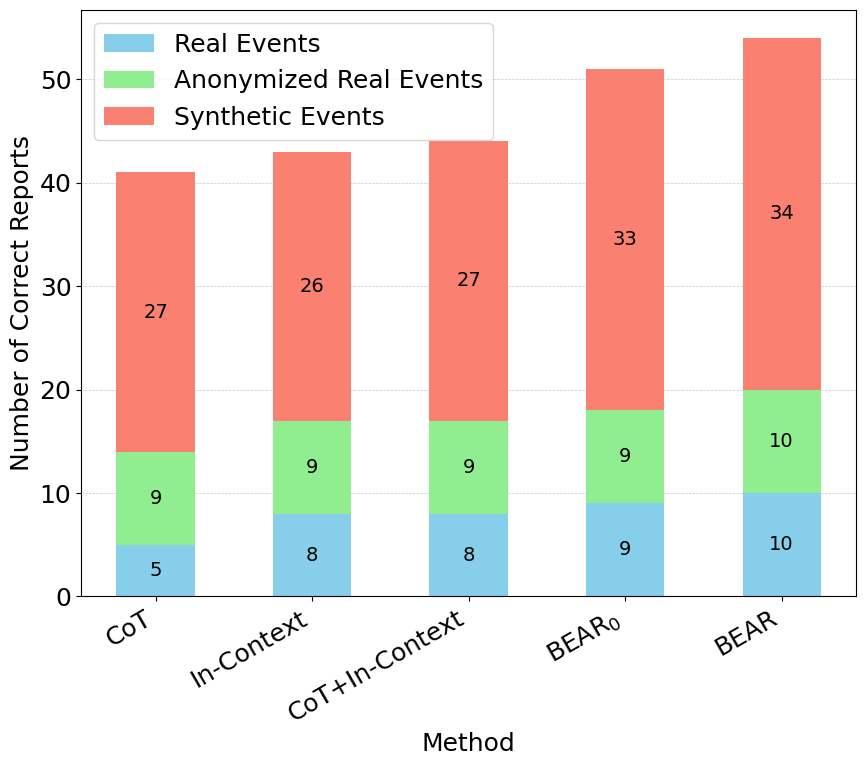}
    \caption{Number of correct reports generated by BEAR and other naive benchmarks in explaining BGP anomaly events, evaluated on 10 real events, 10 anonymized real events, and 34 synthetic events.}
    \label{fig:result}
\end{figure}

We evaluate our proposed method on both real and synthetic datasets, comprising a total of 54 events. In our experiments, we set $N=5$ in self-consistency and use GPT-4o as the backbone LLM. Additionally, we evaluate BEAR on two other LLMs (Claude-3.7-Sonnet and Llama-3.3-70B-Instruct) available through the Amazon Bedrock service, with the corresponding results presented at the end of this section.

We compare BEAR against several baselines, including chain-of-thought (CoT) reasoning \cite{wei2022chain}, in-context learning \cite{dong2024survey}, and a combination of both (CoT+In-Context), as these are the most common prompt engineering methods in natural language processing. In the CoT reasoning baseline, we provide definitions for both BGP hijack and BGP route leak, instructing the LLM to explain its reasoning when inferring the event type, thereby encouraging a step-by-step thought process. For the in-context learning baseline, we present the LLM with four synthetic examples of BGP data generated by our synthetic BGP event generation framework. Each example corresponds to a specific event type—namely, one BGP hijack, one BGP sub-prefix hijack, one BGP route leak, and one BGP sub-prefix route leak—and is accompanied by its respective event type and explanation. Notably, these examples are not included in the synthetic dataset. The hybrid approach, CoT+In-Context, combines both strategies by including the definitions and example reports in the prompt. Additionally, we assess the performance of BEAR without the self-consistency mechanism, referred to as BEAR$_0$. All generated reports are reviewed by BGP experts for evaluation.

\begin{figure}
    \centering
    \includegraphics[width=0.9\linewidth]{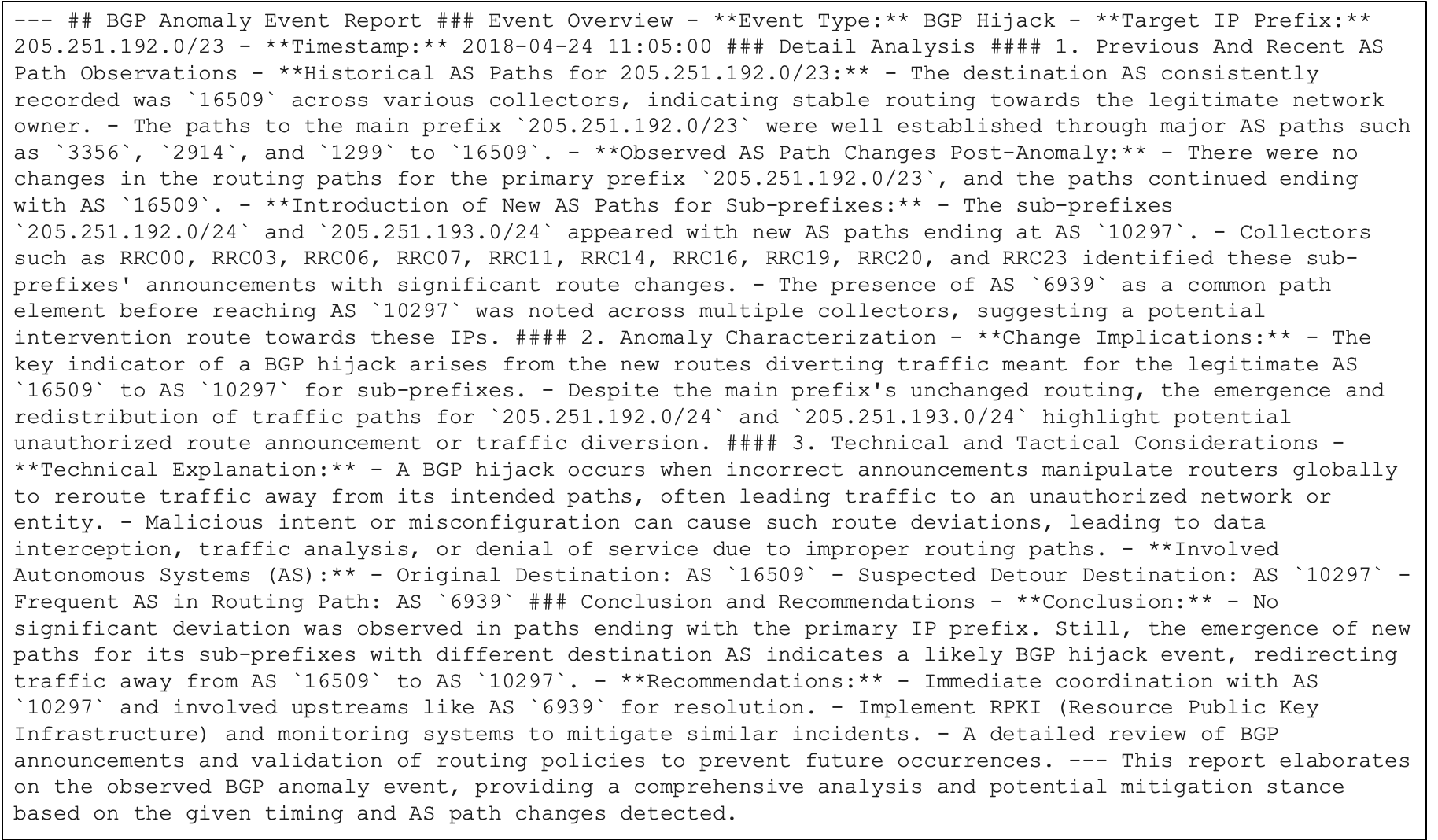}
    \caption{An example report for a BGP hijack event. In this report, BEAR not only identifies the event type but also detects the hijacked sub-prefix.}
    \label{fig:report_exp}
\end{figure}
As shown in Figure \ref{fig:result}, BEAR achieves 100\% accuracy in explaining BGP anomaly events, outperforming all baseline methods. BEAR$_0$ achieves 90.7\% accuracy, surpassing CoT, In-Context, and CoT+In-Context, highlighting the significant improvement in data analysis capabilities enabled by the multi-step reasoning approach. Furthermore, the superior performance of BEAR compared to BEAR$_0$ demonstrates that the self-consistency mechanism enhances the accuracy of the generated reports. Figure \ref{fig:report_exp} presents an example report. Despite the increased complexity introduced by the involvement of a sub-prefix, BEAR successfully identifies all key information and generates a comprehensive report. Additionally, the report includes recommended actions to remedy the anomaly.

\begin{figure}
    \centering
    \includegraphics[width=\columnwidth]{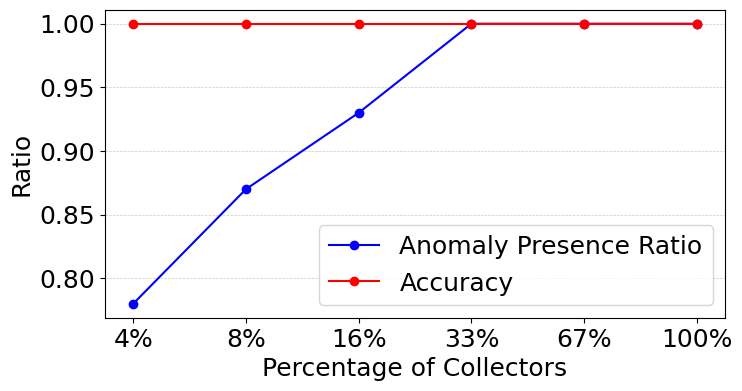}
    \caption{Experimental results under limited data conditions (x-axis shows the percentage of collectors). The anomaly presence ratio represents the percentage of anomaly events captured in $D'_{before}$ and $D'_{after}$. Accuracy accounts for reports that either correctly explain the BGP anomaly event or produce an inconclusive report due to missing event-related data while recommending additional data collection from other collectors.}
    \label{fig:few_collector}
\end{figure}

Alfroy et al. \cite{alfroy2024next} demonstrate that public BGP collection platforms, such as RIPE RIS, capture data from only a limited fraction of ASes. As a result, while historical data offers near-complete routing visibility, real-time anomaly monitoring depends on a restricted set of collectors. Their simulations indicate that only about 16\% of peer-to-peer links are observed in real time, causing roughly 24\% of simulated BGP hijacks to be missed. Moreover, RIPE RIS collectors can become temporarily inaccessible due to connectivity issues or maintenance. For instance, on January 16, 2025, several route collectors experienced connectivity problems that prevented data transmission \cite{ripeout}. Similarly, on February 10, 2025, the rrc19 collector was taken offline for hardware migration, interrupting its BGP sessions until maintenance was concluded \cite{ripePlannedMaintenance}. Motivated by these findings, we evaluate BEAR’s performance under conditions of partial data availability during a BGP anomaly, reflecting real-world scenarios in which comprehensive BGP data may be inaccessible. To simulate this limitation, we randomly select a subset of collectors from the 24 RIPE RIS collectors and restrict the BGP data in $D_{before}$ and $D_{after}$ to those selected collectors, resulting in $D'_{before}$ and $D'_{after}$. Dataset $D_{history}$ remains unchanged as a full historical reference for the LLM to identify missing data during report generation.

Our objective is twofold: (1) if the anomaly event is represented in $D'_{before}$ and $D'_{after}$, BEAR should generate an accurate BGP anomaly event report; (2) if the event is absent in $D'_{before}$ and $D'_{after}$, BEAR should produce an inconclusive report and recommend collecting additional data from other collectors, using $D_{history}$ as a guide. 

\begin{figure}
    \centering
    \includegraphics[width=0.9\linewidth]{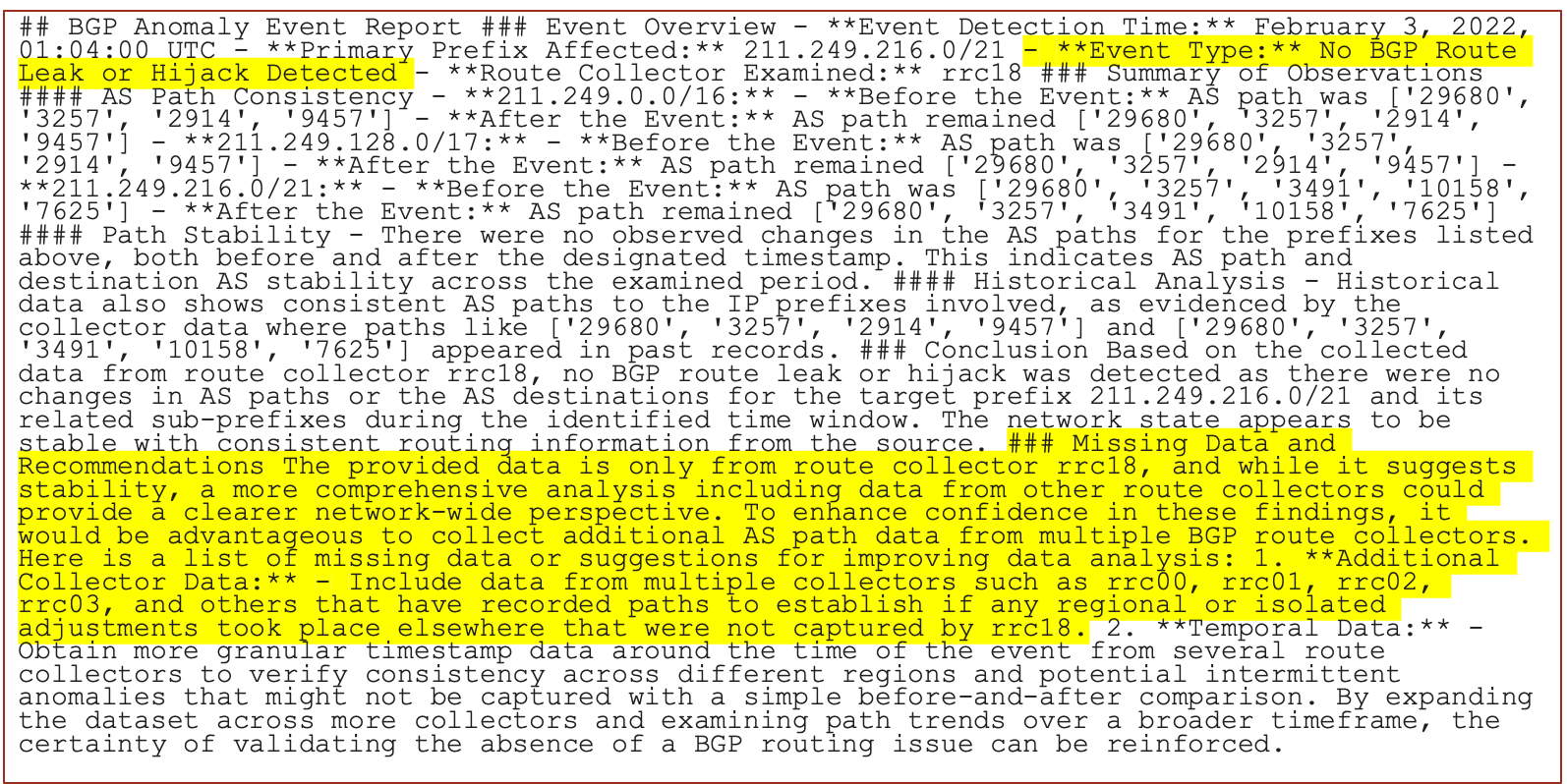}
    \caption{Example report when $D'_{before}$ and $D'_{after}$ does not contain the anomaly event.}
    \label{fig:e_report_partial}
\end{figure}
As shown in Figure \ref{fig:few_collector}, increasing the number of collectors leads to a higher capture rate of anomaly events in the BGP data, as expected, while BEAR maintains consistent 100\% accuracy. In scenarios where data from only a single collector is available (4\% of collectors), 78\% of anomaly events were present in $D'_{before}$ and $D'_{after}$, including 7 real, 2 anonymous real, and 33 synthetic events, and BEAR successfully generated accurate reports for all these cases. For the remaining 22\% of events, where anomaly-related data was missing, BEAR produced inconclusive reports while appropriately recommending additional data collection from other collectors, thereby still fulfilling our predefined objective. With two collectors (8\% of collectors), 87\% of events are captured in $D'_{before}$ and $D'_{after}$, comprising of 7 real, 6 anonymous real, and 34 synthetic events. With four collectors (16\% of collectors), 93\% of anomaly events are presented in $D'_{before}$ and $D'_{after}$, including 8 real, 8 anonymous real, and 34 synthetic events. When 33\% or more of collectors are available, all anomaly events are reflected in $D'_{before}$ and $D'_{after}$. As illustrated in Figure \ref{fig:e_report_partial}, the highlighted part demonstrates BEAR's ability to recognize the lack of necessary event data and suggest gathering more information to complete the analysis.

\begin{figure}
    \centering
    \includegraphics[width=\columnwidth]{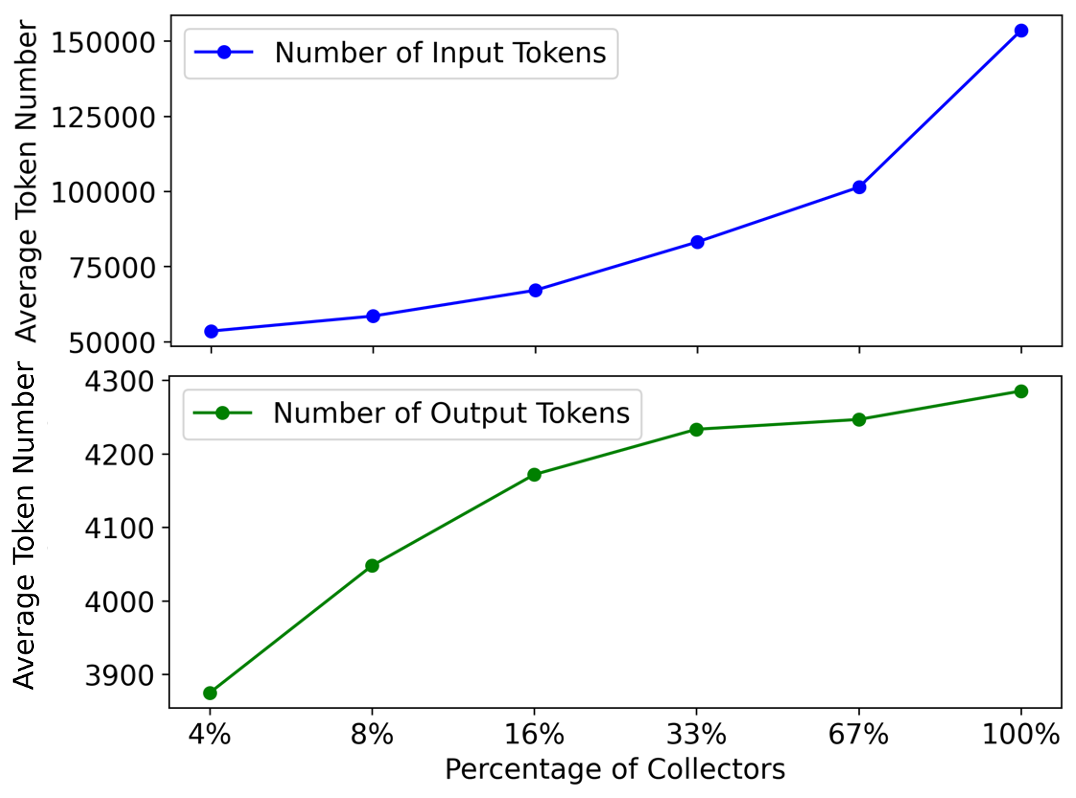}
    \caption{Average number of tokens required to generate a report at different levels of available data. The x-axis represents the percentage of collectors. The two plots show how resource requirements vary as the amount of provided data increases: the first plot displays the number of input tokens, and the second plot shows the number of output tokens.}
    \label{fig:time_genrate}
\end{figure}

Additionally, we evaluate the token resources required for report generation under different levels of collectors, as illustrated in Figure \ref{fig:time_genrate}. The token counts represent the average over all 54 events (both real and synthetic), since resource usage is similar across real and synthetic events. In general, the total token count used by BEAR to generate a BGP anomaly explanation increases sublinearly with the number of collectors—and, consequently, with the data volume. When employing all 24 collectors, BEAR requires an average of 153,628 input tokens and 4,285 output tokens per report. In contrast, when only 4\% of collectors are available (i.e., one collector), BEAR uses an average of 53,482 input tokens and 3,874 output tokens per report.

\begin{figure}
    \centering
    \includegraphics[width=0.9\linewidth]{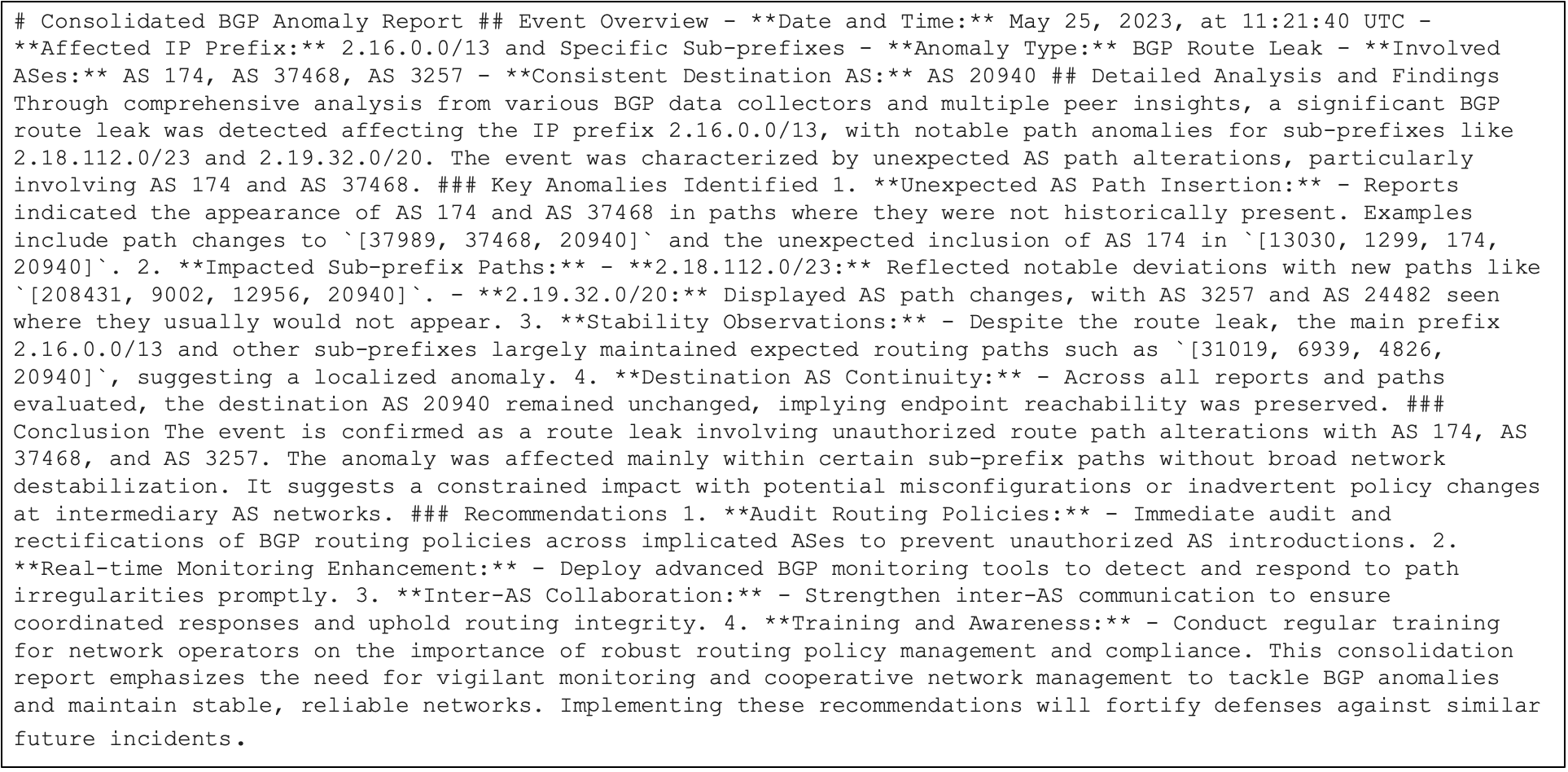}
    \caption{An example report of a BGP anomaly event with an extensive volume of data.}
    \label{fig:big_data_report}
\end{figure}

During data collection, we observe that one BGP anomaly event involves data volumes too large for the backbone LLM to process. Specifically, event $E_e$ — a BGP route leak at Angola Cables \cite{manrsRouteLeak} on May 25, 2023 — contains an unusually high amount of BGP data, with 697 IP prefixes and 771,654 AS paths (in contrast to an average of 2–3 IP prefixes for other events). This high volume results from the target IP prefix having many sub-prefixes, all of which are retrieved to cover the sub-prefix scenario. As this volume exceeds the LLM’s token limit, we excluded $E_e$ from previous experiments.

To address this issue, we develop a hierarchical summarization strategy. When the BGP data is exceptionally large, we partition it either by collector or by peer based on which method minimizes the number of segments while staying within the token limit. Suppose the data is divided into $M$ segments, where a segment is either BGP data of a collector or a peer. BEAR is then applied to each segment to generate $M$ initial reports, $R = \{r_1, r_2, \ldots, r_M\}$, where each report $r_i$ is generated from the BGP data of the $i$th segment. These reports are progressively summarized in batches of $x$ reports (with $x \ll M$). Specifically, the backbone LLM summarizes each batch of $x$ reports into a single report, yielding a new set $R^1 = \{r^1_1, r^1_2, \ldots, r^1_N\}$, where $N = \lceil M/x \rceil$. To obtain the summarized report $r^1_j$, the backbone LLM is instructed to summarize $x$ random reports from $R$ with a focus on information relevant to the BGP anomaly. This summarization is iterated $k$ times until only one final report remains in $R^k$, which serves as the comprehensive explanation of the anomaly.

We test this method on $E_e$ and its anonymous counterpart with $x = 5$. In both cases, partitioning by peer (yielding $M = 576$) is necessary, as partitioning by collector does not sufficiently reduce the data size and still exceeds the token limit. The summarization process is repeated $k = 4$ times to produce the final report. The final reports generated using this strategy are accurate. However, high computational costs limit our experiments to $E_e$ and its anonymous counterpart. One such report is presented in Figure \ref{fig:big_data_report}.

\begin{figure}
    \centering
    \includegraphics[width=\columnwidth]{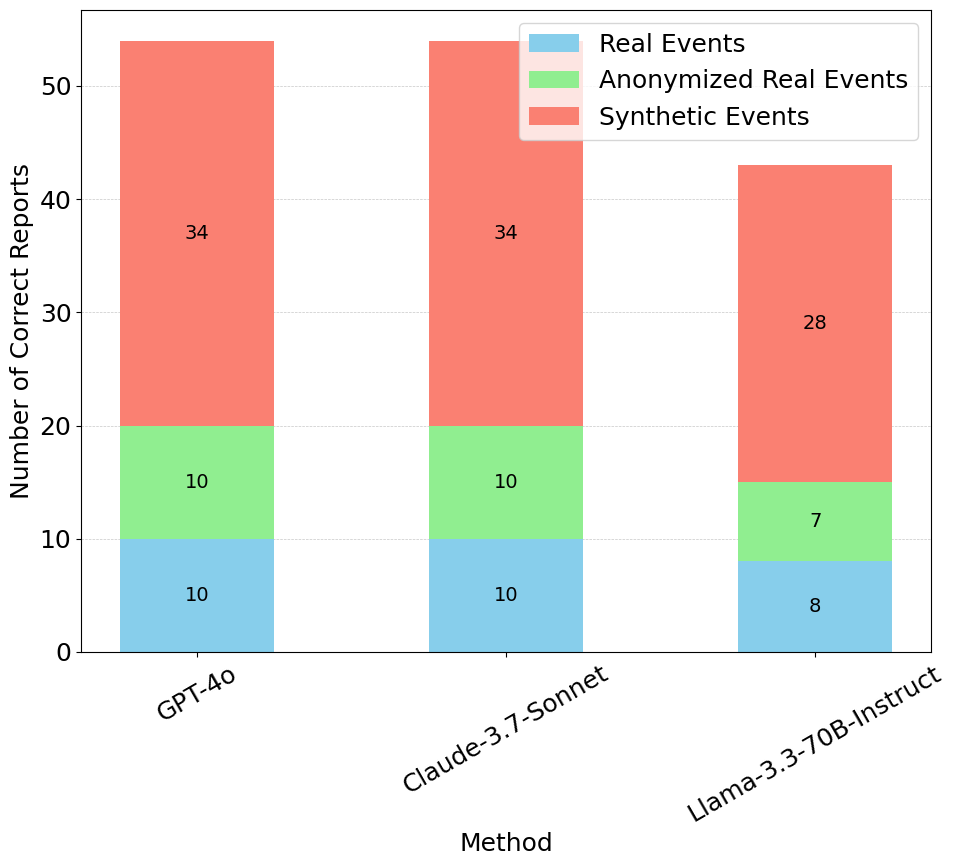}
    \caption{Performance of BEAR using different LLM as backbone model. Evaluated on 10 real events,10 anonymized real events, and 34 synthetic events.}
    \label{fig:diff_LLM}
\end{figure}

Additionally, we assess BEAR using different large language models, with the results depicted in Figure \ref{fig:diff_LLM}. When employing Claude-3.7-Sonnet, BEAR attains an accuracy of 100\% demonstrating performance comparable to BEAR utilizing GPT-4o. In contrast, BEAR achieves only 80\% accuracy with Llama-3.3-70B-Instruct, which is lower than the performance observed with GPT-4o but still surpasses the baseline methods implemented with GPT-4o. With Claude-3.7-Sonnet, BEAR processes an average of 151,923 input tokens and 3,065 output tokens per BGP report. In the case of Llama-3.3-70B-Instruct, BEAR processes an average of 110,272 input tokens and 4,338 output tokens per report. Since BEAR depends on the reasoning capabilities of LLMs, those that excel in this area are more appropriate for explaining BGP anomalies. As demonstrated in Figure \ref{fig:diff_LLM}, employing an LLM with reasoning abilities at least comparable to GPT-4o and Claude-3.7-Sonnet is essential for ensuring the accuracy of BGP report generation. Moreover, in practical deployments, a trade-off between accuracy and cost should be considered.

\section{Conclusion and Future Directions}
\label{sec:conclu}
This paper presents BEAR, a novel framework that addresses the critical challenge of explaining BGP anomaly events. By leveraging the capabilities of LLMs, BEAR transforms raw BGP data into detailed, accurate reports that provide valuable insights for network operators. Our framework's multi-step reasoning approach and integration of self-consistency mechanisms enable high accuracy in anomaly event explanation, even when data availability is limited. Furthermore, this paper presents a synthetic data generation framework that expands evaluation possibilities, mitigating the scarcity of real-world datasets. Experiments demonstrate BEAR's superior performance compared to naive methods, achieving 100\% accuracy on both real and synthetic datasets. This work not only advances the state of BGP anomaly detection and explanation but also lays the foundation for integrating LLMs into broader network management applications. 

\begin{figure}
    \centering
    \includegraphics[width=0.9\linewidth]{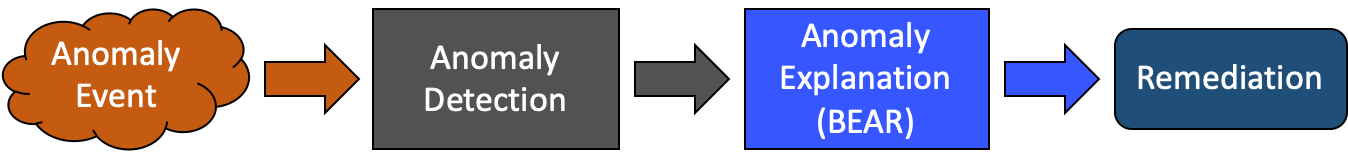}
    \caption{A diagram to the network operations pipeline.}
    \label{fig:future_pipeline}
\end{figure}
Looking ahead, a promising future direction is the development of a comprehensive network operations pipeline, as in Figure \ref{fig:future_pipeline}, that integrates anomaly detection, BGP explanation, and remediation. In this envisioned pipeline, detected BGP anomalies would be automatically fed into the BEAR framework, which would generate interpretations and recommended mitigations. These outputs would then be vetted by network operators and implemented to minimize disruption.

Additionally, BEAR’s approach could be generalized beyond networking, particularly by extending graph-based reasoning to other domains, such as social network analysis and financial fraud detection.

Finally, future work may focus on optimizing BEAR’s cost-efficiency, refining its prompts to further reduce inference costs, and expanding its applicability to other network protocols and security applications.

\bibliographystyle{IEEEtran}
\bibliography{main}

\begin{thebibliography}{10}
\providecommand{\url}[1]{#1}
\csname url@samestyle\endcsname
\providecommand{\newblock}{\relax}
\providecommand{\bibinfo}[2]{#2}
\providecommand{\BIBentrySTDinterwordspacing}{\spaceskip=0pt\relax}
\providecommand{\BIBentryALTinterwordstretchfactor}{4}
\providecommand{\BIBentryALTinterwordspacing}{\spaceskip=\fontdimen2\font plus
\BIBentryALTinterwordstretchfactor\fontdimen3\font minus \fontdimen4\font\relax}
\providecommand{\BIBforeignlanguage}[2]{{%
\expandafter\ifx\csname l@#1\endcsname\relax
\typeout{** WARNING: IEEEtran.bst: No hyphenation pattern has been}%
\typeout{** loaded for the language `#1'. Using the pattern for}%
\typeout{** the default language instead.}%
\else
\language=\csname l@#1\endcsname
\fi
#2}}
\providecommand{\BIBdecl}{\relax}
\BIBdecl

\bibitem{rekhter2006border}
Y.~Rekhter, T.~Li, and S.~Hares, ``A border gateway protocol 4 ({BGP-4}),'' \emph{No. rfc4271}, 2006.

\bibitem{labovitz1998internet}
C.~Labovitz, G.~R. Malan, and F.~Jahanian, ``Internet routing instability,'' \emph{IEEE/ACM Transactions on Networking}, vol.~6, no.~5, pp. 515--528, 1998.

\bibitem{al2015detecting}
B.~Al-Musawi, P.~Branch, and G.~Armitage, ``Detecting {BGP} instability using recurrence quantification analysis ({RQA}),'' \emph{2015 IEEE 34th International Performance Computing and Communications Conference}, pp. 1--8, 2015.

\bibitem{al2016bgp}
B.~Al{-}Musawi, P.~Branch, and G.~Armitage, ``{BGP} anomaly detection techniques: A survey,'' \emph{IEEE Communications Surveys \& Tutorials}, vol.~19, no.~1, pp. 377--396, 2016.

\bibitem{scott2025bgp}
B.~A. Scott, M.~N. Johnstone, P.~Szewczyk, and S.~Richardson, ``{BGP} anomaly detection as a group dynamics problem,'' \emph{Computer Networks}, vol. 257, p. 110926, 2025.

\bibitem{scott2024matrix}
B.~A. {Scott}, M.~N. Johnstone, P.~Szewczyk, and S.~Richardson, ``Matrix profile data mining for {BGP} anomaly detection,'' \emph{Computer Networks}, vol. 242, p. 110257, 2024.

\bibitem{xu2020bgp}
M.~Xu and X.~Li, ``{BGP} anomaly detection based on automatic feature extraction by neural network,'' \emph{2020 IEEE 5th Information Technology and Mechatronics Engineering Conference}, pp. 46--50, 2020.

\bibitem{hoarau2021suitability}
K.~Hoarau, P.~U. Tournoux, and T.~Razafindralambo, ``Suitability of graph representation for {BGP} anomaly detection,'' \emph{2021 IEEE 46th Conference on Local Computer Networks}, pp. 305--310, 2021.

\bibitem{sunita2024optimal}
M.~Sunita and S.~V. Mallapur, ``Optimal detection of border gateway protocol anomalies with extensive feature set,'' \emph{Multimedia Tools and Applications}, vol.~83, no.~17, pp. 50\,893--50\,919, 2024.

\bibitem{dong2021isp}
Y.~Dong, Q.~Li, R.~O. Sinnott, Y.~Jiang, and S.~Xia, ``{ISP} self-operated {BGP} anomaly detection based on weakly supervised learning,'' \emph{2021 IEEE 29th International Conference on Network Protocols}, pp. 1--11, 2021.

\bibitem{moriano2021using}
P.~Moriano, R.~Hill, and L.~J. Camp, ``Using bursty announcements for detecting {BGP} routing anomalies,'' \emph{Computer Networks}, vol. 188, p. 107835, 2021.

\bibitem{latif2022unveiling}
H.~Latif, J.~Pailliss{\'e}, J.~Yang, A.~Cabellos-Aparicio, and P.~Barlet-Ros, ``Unveiling the potential of graph neural networks for {BGP} anomaly detection,'' \emph{Proceedings of the 1st International Workshop on Graph Neural Networking}, pp. 7--12, 2022.

\bibitem{peng2022multi}
S.~Peng, J.~Nie, X.~Shu, Z.~Ruan, L.~Wang, Y.~Sheng, and Q.~Xuan, ``A multi-view framework for {BGP} anomaly detection via graph attention network,'' \emph{Computer Networks}, vol. 214, 2022.

\bibitem{fezeu2020anomalous}
R.~A. Fezeu and Z.-L. Zhang, ``Anomalous model-driven-telemetry network-stream {BGP} detection,'' \emph{2020 IEEE 28th International Conference on Network Protocols}, pp. 1--6, 2020.

\bibitem{huang2024realtime}
J.~Huang, M.~Odiathevar, A.~Valera, J.~Sahni, M.~Frean, and W.~K. Seah, ``Realtime {BGP} anomaly detection using graph centrality features,'' \emph{International Conference on Advanced Information Networking and Applications}, pp. 222--233, 2024.

\bibitem{zhao2023survey}
W.~X. Zhao, K.~Zhou, J.~Li, T.~Tang, X.~Wang, Y.~Hou, Y.~Min, B.~Zhang, J.~Zhang, Z.~Dong \emph{et~al.}, ``A survey of large language models,'' \emph{arXiv preprint arXiv:2303.18223}, 2023.

\bibitem{zhou2025can}
Z.~Zhou and R.~Yu, ``Can {LLM}s understand time series anomalies?'' \emph{The Thirteenth International Conference on Learning Representations}, 2025.

\bibitem{schlamp2016cair}
J.~Schlamp, M.~W{\"a}hlisch, T.~C. Schmidt, G.~Carle, and E.~W. Biersack, ``Cair: Using formal languages to study routing, leaking, and interception in {BGP},'' \emph{arXiv preprint arXiv:1605.00618}, 2016.

\bibitem{buhler2023oscilloscope}
T.~B{\"u}hler, A.~Milolidakis, R.~Jacob, M.~Chiesa, S.~Vissicchio, and L.~Vanbever, ``Oscilloscope: Detecting {BGP} hijacks in the data plane,'' \emph{arXiv preprint arXiv:2301.12843}, 2023.

\bibitem{routeviews2013university}
\BIBentryALTinterwordspacing
{RouteViews}. (2013) University of oregon routeviews project. [Online]. Available: \url{https://routeviews.org}
\BIBentrySTDinterwordspacing

\bibitem{ripe}
\BIBentryALTinterwordspacing
{RIPE}. (2025) {RIPE}: Routing information service ({RIS}). [Online]. Available: \url{https://www.ripe.net/analyse/internet-measurements/routing-information-service-ris/}
\BIBentrySTDinterwordspacing

\bibitem{mai2008detecting}
J.~Mai, L.~Yuan, and C.-N. Chuah, ``Detecting {BGP} anomalies with wavelet,'' \emph{NOMS 2008-2008 IEEE Network Operations and Management Symposium}, pp. 465--472, 2008.

\bibitem{prakash2009bgp}
B.~A. Prakash, N.~Valler, D.~Andersen, M.~Faloutsos, and C.~Faloutsos, ``{BGP}-lens: Patterns and anomalies in internet routing updates,'' \emph{Proceedings of the 15th ACM SIGKDD International Conference on Knowledge Discovery and Data Mining}, pp. 1315--1324, 2009.

\bibitem{li2005internet}
J.~Li, D.~Dou, Z.~Wu, S.~Kim, and V.~Agarwal, ``An internet routing forensics framework for discovering rules of abnormal {BGP} events,'' \emph{ACM SIGCOMM Computer Communication Review}, vol.~35, no.~5, pp. 55--66, 2005.

\bibitem{de2011anomaly}
I.~O. de~Urbina~Cazenave, E.~K{\"o}{\c{s}}l{\"u}k, and M.~C. Ganiz, ``An anomaly detection framework for {BGP},'' \emph{2011 International Symposium on Innovations in Intelligent Systems and Applications}, pp. 107--111, 2011.

\bibitem{al2012machine}
N.~M. Al-Rousan and L.~Trajkovi{\'c}, ``Machine learning models for classification of {BGP} anomalies,'' \emph{2012 IEEE 13th International Conference on High Performance Switching and Routing}, pp. 103--108, 2012.

\bibitem{lutu2014separating}
A.~Lutu, M.~Bagnulo, J.~Cid-Sueiro, and O.~Maennel, ``Separating wheat from chaff: Winnowing unintended prefixes using machine learning,'' \emph{IEEE INFOCOM 2014-IEEE Conference on Computer Communications}, pp. 943--951, 2014.

\bibitem{huang2007diagnosing}
Y.~Huang, N.~Feamster, A.~Lakhina, and J.~Xu, ``Diagnosing network disruptions with network-wide analysis,'' \emph{ACM SIGMETRICS Performance Evaluation Review}, vol.~35, no.~1, pp. 61--72, 2007.

\bibitem{deshpande2009online}
S.~Deshpande, M.~Thottan, T.~K. Ho, and B.~Sikdar, ``An online mechanism for {BGP} instability detection and analysis,'' \emph{IEEE Transactions on Computers}, vol.~58, no.~11, pp. 1470--1484, 2009.

\bibitem{ganiz2006detection}
M.~C. Ganiz, S.~Kanitkar, M.~C. Chuah, and W.~M. Pottenger, ``Detection of interdomain routing anomalies based on higher-order path analysis,'' \emph{Sixth International Conference on Data Mining}, pp. 874--879, 2006.

\bibitem{theodoridis2012novel}
G.~Theodoridis, O.~Tsigkas, and D.~Tzovaras, ``A novel unsupervised method for securing {BGP} against routing hijacks,'' \emph{Computer and Information Sciences III: 27th International Symposium on Computer and Information Sciences}, pp. 21--29, 2012.

\bibitem{karlin2006pretty}
J.~Karlin, S.~Forrest, and J.~Rexford, ``Pretty good {BGP}: Improving {BGP} by cautiously adopting routes,'' \emph{Proceedings of the 2006 IEEE International Conference on Network Protocols}, pp. 290--299, 2006.

\bibitem{lad2006phas}
M.~Lad, D.~Massey, D.~Pei, Y.~Wu, B.~Zhang, and L.~Zhang, ``{PHAS}: A prefix hijack alert system.'' \emph{USENIX Security Symposium}, vol.~1, no.~2, p.~3, 2006.

\bibitem{haeberlen2009netreview}
A.~Haeberlen, I.~C. Avramopoulos, J.~Rexford, and P.~Druschel, ``{NetReview}: Detecting when interdomain routing goes wrong.'' \emph{USENIX Symposium on Networked Systems Design and Implementation}, vol. 2009, pp. 437--452, 2009.

\bibitem{shi2012detecting}
X.~Shi, Y.~Xiang, Z.~Wang, X.~Yin, and J.~Wu, ``Detecting prefix hijackings in the internet with argus,'' \emph{Proceedings of the 2012 Internet Measurement Conference}, pp. 15--28, 2012.

\bibitem{zheng2007light}
C.~Zheng, L.~Ji, D.~Pei, J.~Wang, and P.~Francis, ``A light-weight distributed scheme for detecting {IP} prefix hijacks in real-time,'' \emph{ACM SIGCOMM Computer Communication Review}, vol.~37, no.~4, pp. 277--288, 2007.

\bibitem{hu2007accurate}
X.~Hu and Z.~M. Mao, ``Accurate real-time identification of ip prefix hijacking,'' \emph{2007 IEEE Symposium on Security and Privacy}, pp. 3--17, 2007.

\bibitem{tahara2008method}
M.~Tahara, N.~Tateishi, T.~Oimatsu, and S.~Majima, ``A method to detect prefix hijacking by using ping tests,'' \emph{Proceedings of the 11th Asia-Pacific Symposium on Network Operations and Management: Challenges for Next Generation Network Operations and Service Management}, pp. 390--398, 2008.

\bibitem{zhang2008ispy}
Z.~Zhang, Y.~Zhang, Y.~C. Hu, Z.~M. Mao, and R.~Bush, ``ispy: Detecting ip prefix hijacking on my own,'' \emph{Proceedings of the ACM SIGCOMM 2008 Conference on Data Communication}, pp. 327--338, 2008.

\bibitem{mondal2023llms}
R.~Mondal, A.~Tang, R.~Beckett, T.~Millstein, and G.~Varghese, ``What do {LLM}s need to synthesize correct router configurations?'' \emph{Proceedings of the 22nd ACM Workshop on Hot Topics in Networks}, pp. 189--195, 2023.

\bibitem{kan2024mobile}
K.~B. Kan, H.~Mun, G.~Cao, and Y.~Lee, ``Mobile-llama: Instruction fine-tuning open-source llm for network analysis in 5g networks,'' \emph{IEEE Network}, 2024.

\bibitem{palmero2024providing}
M.~Palmero, K.~P. Annamalai, H.~Singaravelan, D.~Zacks, and J.~W. Capobianco, ``Providing an ai-enabled network assistant for command line interface environments,'' \emph{Technical Disclosure Commons}, 2024.

\bibitem{orsini16bgpstream}
C.~Orsini, A.~King, D.~Giordano, V.~Giotsas, and A.~Dainotti, ``{BGPStream}: A software framework for live and historical {BGP} data analysis,'' \emph{Proceedings of the 2016 Internet Measurement Conference}, p. 429–444, 2016.

\bibitem{wei2022chain}
J.~Wei, X.~Wang, D.~Schuurmans, M.~Bosma, F.~Xia, E.~Chi, Q.~V. Le, D.~Zhou \emph{et~al.}, ``Chain-of-thought prompting elicits reasoning in large language models,'' \emph{Advances in Neural Information Processing Systems}, vol.~35, pp. 24\,824--24\,837, 2022.

\bibitem{dong2024survey}
Q.~Dong, L.~Li, D.~Dai, C.~Zheng, J.~Ma, R.~Li, H.~Xia, J.~Xu, Z.~Wu, B.~Chang \emph{et~al.}, ``A survey on in-context learning,'' \emph{Proceedings of the 2024 Conference on Empirical Methods in Natural Language Processing}, pp. 1107--1128, 2024.

\bibitem{alfroy2024next}
T.~Alfroy, T.~Holterbach, T.~Krenc, K.~Claffy, and C.~Pelsser, ``The next generation of {BGP} data collection platforms,'' \emph{Proceedings of the ACM SIGCOMM 2024 Conference}, pp. 794--812, 2024.

\bibitem{ripeout}
\BIBentryALTinterwordspacing
{RIPE}. (2025) {D}elayed {R}{I}{S} dumps for some route collectors incident report for {R}{I}{P}{E} {N}{C}{C}. [Online]. Available: \url{https://status.ripe.net/incidents/w76jvvvpkqts}
\BIBentrySTDinterwordspacing

\bibitem{ripePlannedMaintenance}
\BIBentryALTinterwordspacing
{R}{I}{P}{E}. (2025) {P}lanned maintenance: {R}{R}{C}19 {N}{A}{P} {A}frica {J}ohannesburg hardware migration. [Online]. Available: \url{https://status.ripe.net/incidents/w76jvvvpkqts}
\BIBentrySTDinterwordspacing

\bibitem{manrsRouteLeak}
\BIBentryALTinterwordspacing
A.~Siddiqui. (2023) {B}{G}{P} route leak at {A}ngola {C}ables slows connectivity for many {A}ustralians. [Online]. Available: \url{https://manrs.org/2023/05/bgp-route-leak-at-angola-cables-slows-connectivity-for-many-australians/}
\BIBentrySTDinterwordspacing

\end{thebibliography}
\vspace{12pt}
\color{red}

\end{document}